\documentclass[9pt,sigplan,table]{ieeetran}
\usepackage[utf8]{inputenc}
\usepackage{amsfonts}
\usepackage{amsmath}
\usepackage{amsbsy}
\usepackage{mathtools}
\usepackage{subfig}
\usepackage{arydshln}
\usepackage[mathscr]{euscript}
\usepackage{graphicx}
\usepackage{xcolor}

\newtheorem{remark}{Remark}

\definecolor{matlabpurple}{rgb}{.627,.126,.941}
\usepackage[framed,numbered,autolinebreaks,useliterate]{mcode}
\usepackage{listings}
\usepackage{mdframed}
\usepackage{url}

\usepackage[strict]{changepage}

\usepackage{framed}

\definecolor{formalshade}{rgb}{0.975,0.975,.975}

\newcounter{main}
\lstnewenvironment{ExampleCode}[1][firstnumber = \themain, name=main]
{\lstset{
			basicstyle={\lstbasicfont},                 % set font
			showstringspaces=false,                     % do not emphasize spaces in strings
			tabsize=4,                                  % number of spaces of a TAB
			mathescape=false,escapechar=§,              % escape to latex with §...§
			upquote=true,                               % upright quotes
			aboveskip=0pt,
			belowskip=0pt, % Set spacing above and below listings
			columns=fixed,                               % nice spacing
			numbers=left,
			xleftmargin=9pt,
			numbersep=5pt,
			frame = none,
			firstnumber=\themain,
			#1
}
}{\setcounter{main}{\value{lstnumber}}}
\newenvironment{fancyequations}{%
	\MakeFramed{\advance\hsize-\width\FrameRestore}%
	\noindent\hspace{-4.55pt}% disable indenting first paragraph
	\begin{adjustwidth}{}{7pt}%
		\vspace{0pt}\vspace{0pt}%
	}
	{%
		\vspace{0pt}\end{adjustwidth}\endMakeFramed\noindent%
}
\usepackage{multirow}

\usepackage{xspace}

\newcommand{\until}{\mathbin{\sf U}}

\newcommand{\X}{\mathbb{X}}
\newcommand{\U}{\mathbb{U}}
\newcommand{\Y}{\mathbb{Y}}

\newcommand{\Bel}{{\mathbf {T}}}

\newcommand{\stochy}{\texttt{StocHy}\xspace}
\newcommand{\syscore}{\texttt{SySCoRe}\xspace}
\newcommand{\amytiss}{\texttt{AMYTISS}\xspace}
\newcommand{\faust}{\texttt{FAUST}\xspace}

\newcommand{\sreachtools}{\texttt{SReachTools}\xspace}
\newcommand{\hypeg}{\texttt{HYPEG}\xspace}
\newcommand{\hpnmg}{\texttt{hpnmg}\xspace}

\newcommand{\modest}{\texttt{Modest\ Toolset}\xspace}
\newcommand{\mascot}{\texttt{Mascot-SDS}\xspace}

\newcommand{\probreach}{\texttt{ProbReach}\xspace}

\newenvironment{Example}{\itshape}

\begin{document}

\title{\syscore: Synthesis via Stochastic Coupling Relations$^*$}
\author{\thanks{$^*$This work is supported by
		the Dutch NWO Veni project CODEC (project number 18244),
		the UK EPSRC New Investigator Award CodeCPS (EP/V043676/1),
		and the Horizon Europe EIC project SymAware (101070802).
	}Birgit van Huijgevoort$\dagger$\thanks{$\dagger$Birgit van Huijgevoort and Sofie Haesaert are with the Eindhoven University of Technology, The Netherlands}, Oliver Sch\"{o}n$\ddagger$\thanks{$\ddagger$Oliver Sch\"{o}n and sadegh Soudjani are with Newcastle University, United Kingdom}, Sadegh Soudjani$\ddagger$, and Sofie Haesaert$\dagger$}
\maketitle

\begin{abstract}
	We present \syscore,
	a \texttt{MATLAB} toolbox that synthesizes controllers for stochastic continuous-state systems to satisfy temporal logic specifications.
	Starting from a system description and a co-safe temporal logic specification, \syscore provides all necessary functions for synthesizing a robust controller and quantifying the associated formal robustness guarantees.
	It distinguishes itself from other available tools by supporting nonlinear dynamics, complex co-safe temporal logic specifications over infinite horizons and model-order reduction.
	To achieve this, \syscore generates a finite-state abstraction of the provided model and performs probabilistic model checking.
	Then, it establishes a probabilistic coupling to the original stochastic system encoded in an approximate simulation relation, based on which a lower bound on the satisfaction probability is computed.
	\syscore provides non-trivial lower bounds for infinite-horizon properties and unbounded disturbances since its computed error does not grow linearly in the horizon of the specification.
	It exploits a tensor representation to facilitate the efficient computation of transition probabilities.
	We showcase these features on several benchmarks and compare the performance of the tool with existing tools.
\end{abstract}

\section{Introduction}
The design of provably correct controllers is crucial for the development of safety-critical systems such as autonomous vehicles and smart energy grids  \cite{lee2016introduction,alur2015principles}.
 To this end, methods for synthesizing controllers for dynamical systems that are guaranteed to satisfy temporal logic specifications have gained an increasing amount of attention in the control community \cite{BK08,tabuada2009verification,belta2017formal,lavaei2022automated}. Besides establishing the theory underlying these methods, it is equally important to develop tools that facilitate their application.
 For stochastic systems, a collection of tools that can perform formal controller synthesis is already available. A subset of these tools include in alphabetical order:
 \amytiss \cite{lavaei2020amytiss},
 \faust \cite{soudjani2015fau},
 \hpnmg \cite{huls2020hpmngtool},
 \hypeg \cite{pilch2017hypeg},
 \mascot \cite{MMS2020},
 the \modest \cite{hartmanns2014modest},
 \probreach \cite{ProbReach},
 \sreachtools \cite{vinod2019sreachtools},
 and \stochy \cite{cauchi2019stochy}.
 A complete list of these tools with their descriptions and capabilities can be found in the ARCH Competition Report (stochastic category) \cite{ARCH21}.
 These tools perform the computations either using analytical methods or employing statistical model checking.
The approaches in the analytical methods can further be divided into abstraction-based \cite{soudjani2015fau, cauchi2019stochy, lavaei2020amytiss,MMS2020} and abstraction-free techniques \cite{vinod2019sreachtools,kochdumper2021aroc}.
Abstraction-free techniques are generally less prone to suffering from the curse of dimensionality, however, they are often limited to simple invariance and reachability specifications. In contrast, abstraction-based tools can be applied to a breath of systems and specifications. A survey on formal verification and control synthesis of stochastic systems is given in \cite{lavaei2022automated}.

\syscore contributes to the category of tools that employ analytical abstraction-based methods.
It is a \texttt{MATLAB} toolbox 
applicable to stochastic nonlinear systems with a possibly \emph{unbounded disturbance}. Furthermore, it can perform the controller synthesis to satisfy arbitrary co-safe specifications that can have \emph{unbounded time horizons}. To this end, it uses the $(\epsilon,\delta)$-approximate simulation relation provided in \cite{haesaert2017verification}, that explicitly designs the coupling between the continuous-state model and its (reduced) finite-state abstraction \cite{huijgevoort2022similarity}.
Hence, \syscore extends the capabilities of the current tools by considering properties that are \emph{unbounded in time} and by considering systems with an \emph{unbounded disturbance}. 

\syscore is a comprehensive toolbox for temporal logic control of stochastic continuous-state systems, implementing all necessary steps in the control synthesis process. Moreover, it supports \emph{model-order reduction} in the abstraction process with formal error quantification quarantees, which makes it applicable to a larger classes of systems. To increase its computational efficiency, \syscore performs computations based on tensors and sparse matrices. Furthermore, computations based on efficient convex optimizations for polytopic sets are implemented where possible.
The tool is developed with a focus on ease of use and extensibility, such that it can easily be adapted to suit individual research purposes.
The development of \syscore is a step towards solving the tooling need for temporal logic control of stochastic systems as it expands both the class of models and the class of specifications for which abstraction-based methods can provide controllers with formal guarantees.

This tool paper is organized as follows. We discuss in Section~\ref{sec:problem} the temporal logic control problem and the set-up in \syscore. We then give an overview of \syscore in Section \ref{sec:overview} by introducing the associated functions and classes. Section \ref{sec:benchmarks} discusses multiple benchmarks that show the capabilities of \syscore and how it compares to existing tools. We end the paper with a summary and a discussion of possible extensions. 
Throughout, we give the core functions of \syscore in framed white boxes and example code in gray boxes. 

\section{Temporal logic control}
\label{sec:problem}
The main purpose of \syscore is to perform the complete control synthesis procedure in abstraction-based temporal logic control. It is applicable to discrete-time models with a possibly unbounded stochastic disturbance and synthesizes a controller for satisfying co-safe linear temporal logic specifications that may have an unbounded time horizon. The computational approach is based on the theory of approximate simulation relations \cite{haesaert2017verification}, the coupling between models \cite{haesaert2017verification,huijgevoort2022similarity} and robust dynamic programming mappings \cite{haesaert2020robust}. In this section, we introduce the class of models and specifications handled by \syscore, and show how to set up the problem. Furthermore, we provide a high-level description of the theory underlying the implementations in \syscore.
\subsection{Problem parameters}
\textbf{Model.}
%%% Nolinear systems
We consider discrete-time systems described by stochastic difference equations
\begin{equation}
M: \begin{cases}
	x_{t+1} = f(x_t,u_t) + B_w w_t \\
	y_t = Cx_t, \quad \forall t \in \left\{0,1,2, \dots\right\},
\end{cases}
\label{eq:model}
\end{equation}
with state $x_t \in \mathbb{X}$, input $u_t \in \mathbb{U}$, (unbounded) stochastic disturbance $w_t \in \mathbb{W}$, measurable function $f: \mathbb{X}\times \mathbb{U} \rightarrow \mathbb{X}$,  and matrices $B_w$ and $C$ of appropriate sizes.

To handle nonlinear systems of the form \eqref{eq:model} we perform a piecewise-affine (PWA) approximation 
that yields a system described by
\begin{equation}
	\begin{cases}
		x_{t+1} = A_ix_t + B_iu_t + a_i + B_{w,i} w_t + \kappa_t  \text{ for } x_t \in P_i\\
		y_t = Cx_t,
	\end{cases}
	\label{eq:modelPWA}
\end{equation}
with $P_i$ a partition of $\mathbb{X}$ and $\kappa_t \in \mathscr{K}_i$ the error introduced by performing the PWA approximation. For ease of notation, we denote the state-dependent error $\kappa_{x_t}$ as $\kappa_t$. Furthermore, $A_i, B_i, B_{w,i}$ and $a_i$ are matrices of appropriate sizes. Details of temporal logic control for nonlinear stochastic systems via piecewise-affine abstractions can be found in \cite{huijgevoort2022piecewiseaffineabstraction}.
Besides nonlinear systems, we also consider the special case of
linear time-invariant (LTI) systems:
\begin{equation}
\begin{cases}
		x_{t+1} = Ax_t + Bu_t + B_w w_t \\
		y_t = C x_t,
\end{cases}
\label{eq:modelLTI}
\end{equation} with $A$ and $B$ matrices of appropriate sizes.

\begin{remark}
This first release of \syscore assumes the disturbance $w_t$ has unbounded Gaussian distribution $w_t \sim \mathcal{N}(0,I)$. The implementation for other classes of distributions is under way and will be included in the future release of the tool. Note that the assumption of standard Gaussian distribution with zero mean and identity covariance matrix is without loss of generality since any system \eqref{eq:model}-\eqref{eq:modelLTI} with disturbance $w \sim \mathcal{N}(\mu,\Sigma)$ can be rewritten to a system in the same class with an additional affine term \cite{allen2008construction}.
\end{remark}

To specify the model, that is  
a nonlinear system \eqref{eq:model}, a PWA system \eqref{eq:modelPWA} or an LTI system \eqref{eq:modelLTI}, we have developed the classes \texttt{NonLinModel}, \texttt{PWAmodel}, and \texttt{LinModel}, respectively.
The state space, input space, and the sets needed for defining the specification should be defined in these class descriptions.

\subsubsection*{Running example}
\begin{Example}
Consider a two-dimensional (2D) case study of parking a car with dynamics of the form \eqref{eq:modelLTI} with $A = 0.9I_2$, \mbox{$B=0.7I_2$}, and $B_w =C = I_2$. Furthermore, we have state space $\mathbb{X}=[-10,10]^2$, input space $\mathbb{U} =[-1,1]^2$, and disturbance $w \sim \mathcal{N}(0,I_2)$. After specifying the matrices $A,B,C, B_w$, and setting the values for the disturbance $w$ with mean \texttt{mu} and covariance matrix \texttt{sigma} equal to zero and identity respectively,
we can initialize a model in \syscore as follows:
\end{Example}
\begin{fancyequations}
\begin{ExampleCode}
% Set up an LTI model
sysLTI = LinModel(A,B,C,[],Bw,mu,sigma);
\end{ExampleCode}
\end{fancyequations}
\begin{Example}
%\noindent
The state and input spaces are defined using \texttt{Polyhedron} from the \texttt{multi-parametric toolbox (MPT3)} \cite{herceg2013multi} as follows.
\end{Example}
\begin{fancyequations}
\begin{ExampleCode}
% Define bounded state space
sysLTI.X = Polyhedron(combvec([-10,10],[-10,10])');
% Define bounded input space
sysLTI.U = Polyhedron(combvec([-1,1],[-1,1])');
\end{ExampleCode}
\end{fancyequations}

\smallskip\noindent \textbf{Specifications.}
In \syscore, we consider formal specification written using co-safe linear temporal logic (scLTL) \cite{belta2017formal,kupferman2001model}, which consists of atomic proposition (AP) $AP = \left\{ p_1, p_2, \dots, p_N \right\}$ that are either true or false. 
To connect the system and the specification, we label the output space of the system, such that we can relate the trajectories of the system $\boldsymbol{y} = y_0, y_1, y_2, \dots$ to the atomic propositions of the specification $\phi$.
\subsubsection*{Running example cont'd}
\begin{Example}
For the 2D car park, we consider reach-avoid specification $\phi_{park}$ with the region to reach $P_1$ and with avoid region $P_2$. First, we define the regions
\end{Example}
\begin{fancyequations}
\begin{ExampleCode}
% Specify regions for the specification
P1 = Polyhedron([4, -4; 4, 0; 10, 0; 10 -4]);
P2 = Polyhedron([4, 0; 4, 4; 10, 4; 10 0]);
\end{ExampleCode}
\end{fancyequations}
\begin{Example}
and add them to the system object:
\end{Example}
\begin{fancyequations}
\begin{ExampleCode}
% Regions that get specific atomic propositions
sysLTI.regions = [P1;P2];
%  Propositions corresponding to the regions
sysLTI.AP = {'p1', 'p2'};
\end{ExampleCode}
\end{fancyequations}
\begin{Example}
Implicitly, this means that states inside regions \texttt{P1} and \texttt{P2} are labeled using the corresponding atomic propositions \lstinline[]|'p1'| and \lstinline[]|'p2'|, respectively. Now, we can write the scLTL specification
	\begin{equation}
		\phi_{park} =  \neg p_2 \until p_1,
		\label{eq:phi_park}
	\end{equation} using the syntax from \cite{gastin2001fast} as follows
\end{Example}
\begin{fancyequations}
\begin{ExampleCode}
% Define the scLTL specification
formula = '(!p2 U p1)';
\end{ExampleCode}
\end{fancyequations}

\medskip

Denote the system $M$ under the controller $C$ by $M\times C$ as \mbox{in \cite{tabuada2009verification}.} The goal is to synthesize a controller $C$, such that the controlled system satisfies an scLTL specification $\phi$, denoted as $M\times C \models \phi$.
Since we consider stochastic systems, we compute the \emph{satisfaction probability} denoted as $\mathbb{P}(M\times C \models \phi)$.  This goal is formulated mathematically next.

\medskip

\noindent\textbf{Problem statement.}
Given model $M$, scLTL specification $\phi$, and probability threshold $\rho\in(0,1)$, design controller $C$ such that 
\begin{equation}
	\mathbb{P}(M\times C \models \phi) \geq \rho.
	\label{eq:ContrProb}
\end{equation}

\syscore automatically synthesizes a controller by maximizing the right-hand side of \eqref{eq:ContrProb} on a simplified abstract model and makes the computations robust with respect to the abstraction errors.  It provides a robust lower bound on the satisfaction probability, which then can be used by the user to compare with the probability threshold $\rho$.
\begin{figure}
	\centering
	\includegraphics[width=\columnwidth]{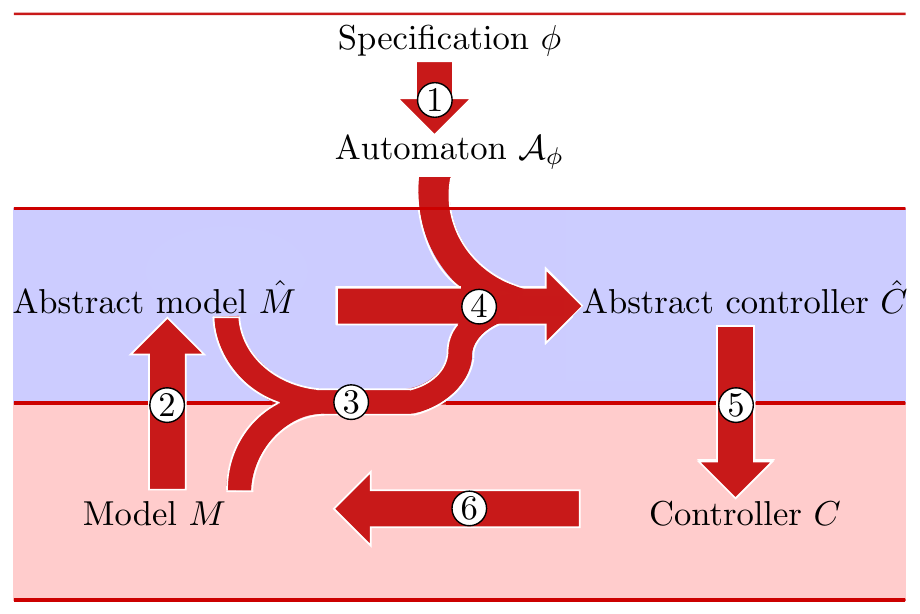}
	\caption{Steps in abstraction-based temporal logic control with 3 main layers: Continuous-state (red), finite-state (blue), and specification (white). The numbers correspond to the following steps: (1) translating the specification to an automaton, (2) (reduced) finite-state abstraction, (3) similarity quantification, (4) synthesizing a controller, (5) control refinement, and (6) deployment.}
	\label{fig:TLcontrol}
\end{figure}
\subsection{Stochastic coupling relations for control synthesis}
To solve the above problem, we use an abstraction-based approach and the \emph{dynamic programming mappings} from \cite{haesaert2020robust}, which allows us to consider infinite-horizon properties.
More specifically, the abstraction-based temporal logic control implemented in \syscore has six main steps, namely (1) translating the specification to an automaton, (2) constructing a (reduced) finite-state abstraction, (3) quantifying the similarity, (4) synthesizing a controller, (5) control refinement, and (6) deployment.

As visualized in Figure~\ref{fig:TLcontrol}, we start from a temporal logic specification that expresses the desired behavior of the controlled system and translate it to an automaton (see top layer). A finite abstract model $\hat{M}$ of the system is also constructed (step 2). For this abstract model $\hat{M}$, its bounded deviation from the original model can be quantified using simulation relations (step 3) \cite{haesaert2017verification,huijgevoort2022similarity}. Computing these bounds is based on an efficient invariant set computation formulated as an optimization problem constrained by a set of parameterized matrix inequalities \cite{huijgevoort2022similarity}. 
Based on the automaton, an abstract controller over the abstract model can be synthesized.
In step 4, we synthesize an abstract controller $\hat{C}$ and compute the robust satisfaction probability. 
The robust satisfaction probability takes the deviation bounds computed in step 3 into account and gives a lower bound on the actual satisfaction probability.

 To compute the robust satisfaction probability and to synthesize an abstract controller $\hat{C}$, \syscore solves a reachability problem over the abstract system combined with the automaton corresponding to the specification. This reachability problem is then solved as a dynamic programming problem.
It is shown in \cite{haesaert2020robust} that leveraging the deviation bounds from step 3, the controller for the abstract model can be refined to the original continuous-state model while preserving the guarantees. 
To construct this controller $C$, \syscore refines the abstract controller in step 5. The resulting controller $C$ is a policy that can be represented with finite memory. Finally,  \syscore deploys the controller on the model (step 6). It is important to note that the abstraction step (step 2 in Figure~\ref{fig:TLcontrol}) can additionally contain model-order reduction or piecewise-affine approximation, which shows the comprehensiveness of \syscore enabled by establishing coupled simulation relations.

The next section gives a complete overview of the toolbox and specifies how each of the steps from Figure~\ref{fig:TLcontrol} is implemented.

\section{Toolbox overview}\label{sec:overview}
After setting-up the problem by specifying the system using the classes \texttt{NonLinModel}, \texttt{PWAmodel} or \texttt{LinModel}, and the specification %either 
as an scLTL formula, we continue with the steps illustrated in Figure~\ref{fig:TLcontrol}. Each step corresponds to a specific function as in Table~\ref{tab:functions}. Note that the abstraction step may have multiple (formal) approximation stages depending on the type of the model or its dimension.

	\begin{center}
		\begin{table}
		\caption{Main functions of \syscore for steps (1)-(6), with optional steps (2a) and (2b).}
			\label{tab:functions}
			\begin{tabular}{ p{3.7cm}  l }
				 \rowcolor{gray!60}
				\hline
				Step & Function  \\ 
				(1) Translate the specification & \texttt{TranslateSpec} \\
				 \rowcolor{gray!25}
				(2) Finite-state abstraction & \texttt{FSabstraction} \\
				 \rowcolor{gray!25}
				(2a) Piecewise-affine approx. & \texttt{PWAapproximation} \\
				 \rowcolor{gray!25}
				(2b) Model-order reduction & \texttt{ModelReduction} \\
				
				(3) Similarity quantification & \texttt{QuantifySim}  \\
				 \rowcolor{gray!25}
				(4) Synthesize a controller & \texttt{SynthesizeRobustController}  \\
				
				(5) Control refinement & \texttt{RefineController}  \\
				 \rowcolor{gray!25}
				(6) Deployment& \texttt{ImplementController} \\
			\end{tabular}
		\end{table}
	\end{center}

\subsection{Translating the specification}
For control synthesis, the scLTL specification is written as a deterministic finite-state automaton (DFA) \cite{belta2017formal}. Examples of such DFAs are given in Figure~\ref{fig:DFA_examples}. 
We use the tool \texttt{LTL2BA}\footnote{Tool available at \url{http://www.lsv.fr/~gastin/ltl2ba/index.php}} to translate an scLTL specification, which constructs a non-deterministic B\"uchi automaton for a general LTL specification \cite{gastin2001fast}. Additionally, we check whether the given formula is written using scLTL (instead of full LTL) and then (if possible) rewrite the non-deterministic B\"uchi automaton to a DFA. This step is based on powerset conversion \cite{rabin1959finite} that is used to convert a nondeterministic finite-state automaton to a DFA. The complete translation from an scLTL specification to a DFA is implemented in the function \texttt{TranslateSpec}.
\begin{lstlisting}[numbers=none]
% Translate an scLTL formula to a DFA
DFA = TranslateSpec(formula, AP);
\end{lstlisting}
The input \texttt{formula} is given using the syntax of \texttt{LTL2BA} in  \cite{gastin2001fast}.
\subsubsection*{Running example cont'd}
\begin{Example}
	For the 2D car park, we consider the reach-avoid specification $\phi_{park}$ in \eqref{eq:phi_park}, which we translate to a DFA using \texttt{TranslateSpec} with \texttt{AP} and \texttt{formula} 
	given respectively in 
	code lines 13 and 15.
\end{Example}
\begin{fancyequations}
\begin{ExampleCode}
% Translate the spec to a DFA
DFA = TranslateSpec(formula, sysLTI.AP);
\end{ExampleCode}
 \end{fancyequations}
Besides reach-avoid specifications it is also possible to describe many other types of specifications, such as more complex reach-avoid specification, e.g. $\phi_{PD} = \lozenge (p_1 \land (\neg p_2 \until p_3))$, or time-bounded and unbounded safety specifications, e.g. $\phi_{BAS} = \bigwedge_{i=0}^{5} \bigcirc^i p_1$ and $\phi_{vdPol} = p_1 \until p_2$. These specifications are written in \syscore as
\begin{subequations} 
\begin{flalign} \SwapAboveDisplaySkip
&\text{\lstinline[language=Matlab]|	formula_PD = 'F(p1 & (!p2 U p3))'; |}&&
	\label{eq:phi_PD} \\ 
&\text{\lstinline[language=Matlab, breaklines=true]|	formula_BAS = '(p1 & X p1 & X X p1 & X X X p1 ... |}&&\label{eq:phi_BAS}  \\
&\text{\lstinline[literate={p1}{{{\color{matlabpurple}p1}}}2 {X}{{{\color{matlabpurple}X}}}2 {\&}{{{\color{matlabpurple}\&}}}2 {p1)'}{{{\color{matlabpurple}p1)'}}}3] |		& X X X X p1 & X X X X X p1)'; |}&& 
\notag \\
&\text{\lstinline[language=Matlab]|	formula_vdPol = '(p1 U p2)'; |}&&
\label{eq:phi_VdP}
\end{flalign}
\end{subequations}

\begin{figure}[t]
	\centering
	\subfloat[DFA corresponding to specification $\phi_{park}$ in \eqref{eq:phi_park} for the running example.]{\includegraphics[width=0.4\columnwidth]{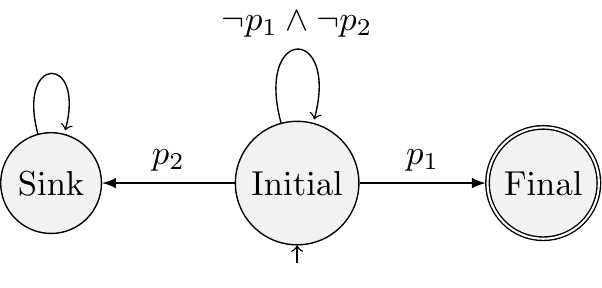}\label{fig:DFA_acyclic}} \quad
	\subfloat[Atypical DFA]{\includegraphics[width=0.55\columnwidth]{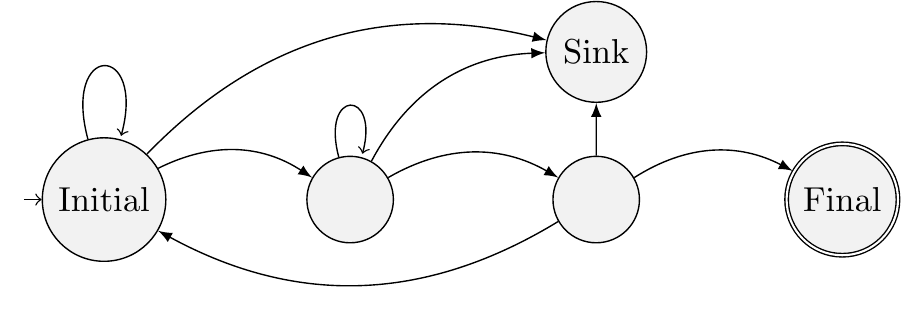}\label{fig:DFA_cyclic}}
	\caption{Acyclic DFA in (a) versus cyclic DFA in (b).}
	\label{fig:DFA_examples}
\end{figure}

Note that it is also possible to directly pass a DFA as an input to \syscore instead of giving the specification as an scLTL formula.
\syscore is able to natively handle both acyclic and cyclic DFAs (see Figure~\ref{fig:DFA_examples}), in contrast to many other tools \cite{soudjani2015fau,rungger2016scots,cauchi2019stochy,lavaei2020amytiss} that do not natively support DFAs but often rely on external tools such as PRISM \cite{kwiatkowska2002prism} to compute the controller. 

\subsection{Abstraction}\label{sec:abstr}
\syscore includes two possible abstraction methods,
namely
finite-state abstraction for 
continuous-state systems in \eqref{eq:model}-\eqref{eq:modelLTI} and
model-order reduction for continuous-state LTI systems \eqref{eq:modelLTI}. However, in order to create a finite-state abstraction of a nonlinear system \eqref{eq:model} we require an additional approximation step before constructing a piecewise-affine finite-state abstraction. Note that the piecewise affine approximation itself is considered as an integral part of the finite-state abstraction method. 

\smallskip
\noindent \textbf{Piecewise affine approximation.}
To approximate a nonlinear system \eqref{eq:model} by a PWA system \eqref{eq:modelPWA}, we partition the state space and use a standard first-order Taylor expansion to approximate the nonlinear dynamics in each partition by affine dynamics. Additionally, we compute the error 
introduced by this approximation. 
In \syscore, this is performed by the function \texttt{PWAapproximation}.
\begin{lstlisting}[numbers=none]
% Perform piecewise-affine approximation
sysPWA = PWAapproximation(sysNonLin, Np);
\end{lstlisting}
Here, the nonlinear system \eqref{eq:model} is given by \texttt{sysNonLin} and the number of partitions in each direction is given by \texttt{Np}. The result is a PWA system \eqref{eq:modelPWA} \texttt{sysPWA}.

\begin{figure}[t]
	\centering
	\subfloat[Coupling between models $M$ and its finite-state abstraction $\hat{M}$ through their inputs and disturbances via  an interface function and a coupling kernel.]{\includegraphics[width=0.4\columnwidth]{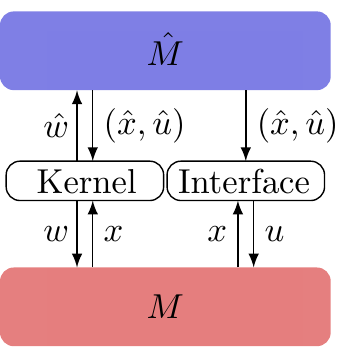}\label{fig:coupling}} \quad
	\subfloat[Coupling between continuous-state models $M$ and $M_r$, and between $M_r$ and its finite-state abstraction $\hat{M}$.]{\includegraphics[width=0.4\columnwidth]{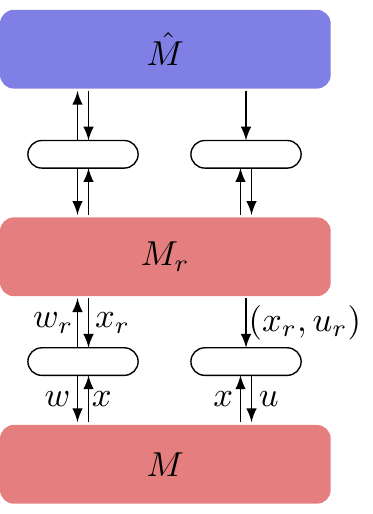}\label{fig:couplingMOR}}
	\caption{Coupling between different models. Red and blue boxes correspond to respectively continuous-state and finite-state. In (a) only a finite-state abstraction is performed, while in (b) both model-order reduction and a finite-state abstraction are shown.}
	\label{fig:CouplingTotal}
\end{figure}

\smallskip
 \noindent \textbf{Interface function.} 
\syscore can construct a reduced-order abstract model $M_r$ and a finite-state abstract model $\hat{M}$ of the original model $M$. Let us denote the control inputs of these models respectively by $u_r$ and $\hat u$. The abstract control inputs $u_r$ and $\hat{u}$ need to be refined to a control input $u$ for $M$ as illustrated in Figure~\ref{fig:CouplingTotal}. 
The input refinement is performed by one or multiple interface functions, namely
\begin{subequations}\label{eq:interface}\begin{align} \SwapAboveDisplaySkip
		& \hspace{-.5cm}
		u_{r,t} =\hat{u}_t
		\hspace{1.5cm} & \mbox{\itshape \small (default) }\label{eq:int1}\\
		& \hspace{-.5cm}
		u_{r,t} =\hat{u}_t + K(x_{r,t}-\hat{x}_t) &
		\mbox{\itshape \small (option 1)} \label{eq:int2} \\
		& \hspace{-.5cm}
			u_t = u_{r,t} + Qx_{r,t}+K_{MOR}(x_t-Px_{r,t}). & 
		\mbox{\itshape \small (option 1, MOR)}
		\label{eq:int_MOR}
\end{align}\end{subequations} 

To refine the input $\hat{u}$ of a finite-state model to the input $u_r$ of a continuous-state reduced-order model, we implemented two 
different interface functions in the format of \eqref{eq:int1} and \eqref{eq:int2}. For many cases the default interface function \eqref{eq:int1} should work fine, however, the option \eqref{eq:int2} gives more influence on the refined controller by including a feedback term. When the interface function \eqref{eq:int2} is used, we have to take this into account when constructing the finite-state abstraction to avoid the input bounds being violated, therefore, the interface function must be chosen before constructing the finite-state abstraction.
We further use the interface function \eqref{eq:int_MOR} to refine the input $u_r$ of a reduced-order model to the input $u$ of the full-order model.
 It should be noted that if only a finite-state abstraction is performed without using model-order reduction (MOR), we have $P=I, Q=0$ and $x_t=x_{r,t}$, hence 
 we obtain interface functions \eqref{eq:int1} and \eqref{eq:int2} with $u_t = u_{r,t}$ and $x_t=x_{r,t}$.
\subsubsection*{Running example cont'd}
\begin{Example} It is required to select an interface function for the input refinement before starting with the temporal logic control steps. 
For this running example only use the default interface function \eqref{eq:int1} without model-order reduction, that is $u_t = \hat{u}_t$. However, if desired, the user can select the option \eqref{eq:int2} by setting \texttt{int\_f = 1} and passing this to the functions.
\end{Example}

In the remainder of this section, we discuss how to obtain the reduced-order and finite-state abstract models.

\smallskip\noindent \textbf{Model-order reduction.} \label{sec:mor}
It is essential to include model-order reduction for high-dimensional models. For LTI systems \eqref{eq:modelLTI} this yields a reduced-order model $M_r$ of the form
\begin{equation}
	M_r:	\begin{cases}
		x_{r,t+1} = A_rx_{r,t} + B_ru_{t} + B_{rw} w_{r,t} \\
		y_{r,t} = C_r x_{r,t},
	\end{cases}
	\label{eq:modelRed}
\end{equation}
with $x_r \in \mathbb{X}_r, u\in \mathbb{U}$, $y \in \mathbb{Y}$, $w_r \in \mathbb{W}$, and matrices $A_r, B_r, B_{rw}$ and $C_r$ of appropriate sizes. 

In \syscore, the function \texttt{ModelReduction} constructs a reduced-order model \texttt{sysLTIr} of dimension \texttt{dimr} based on the original model \texttt{sysLTI} by using \emph{balanced truncations} on a closed loop system with 
a feedback matrix $F$. This feedback matrix is computed by solving \emph{discrete-time algebraic Riccati equations} that can be tuned using constant \texttt{f} \cite{pappas1980numerical}. The syntax of \texttt{ModelReduction} is
\begin{lstlisting}[numbers=none]
% Construct reduced-order model
[sysLTIr, F] = ModelReduction(sysLTI, dimr, f)
\end{lstlisting}

We couple the inputs $u,u_r$ from $M$ \eqref{eq:modelLTI} and $M_r$ \eqref{eq:modelRed} using the interface function \eqref{eq:int_MOR} as illustrated in Figure~\ref{fig:couplingMOR}. This is based on the theoretical results presented in \cite{haesaert2017verification,huijgevoort2022similarity}.
To compute matrices $P$ and $Q$ for the interface function, we have the function \texttt{ComputeProjection} that adds the matrices automatically to the object \texttt{sysLTIr}.
\begin{lstlisting}[numbers=none]
% Compute matrices P and Q
sysLTIr = ComputeProjection(sysLTI, sysLTIr);
\end{lstlisting} 

\smallskip
\noindent\textbf{Finite-state abstraction.}
We grid the state space to construct a finite-state abstraction $\hat{M}$ of the continuous-state models 
\eqref{eq:modelPWA}, \eqref{eq:modelLTI} or \eqref{eq:modelRed}. More specifically, we compute the abstract state space $\mathbb{\hat{X}}$ as the set consisting of the centers of the grid cells. Next, the dynamics of the abstract model is defined by using the operator $\Pi: \mathbb{X} \rightarrow \mathbb{\hat{X}}$ that maps states $x$ to the center of the grid cell it is in. Details on how to construct a finite-state abstraction of a nonlinear system or an LTI system can be found in \cite[Section III]{huijgevoort2022piecewiseaffineabstraction}, and \cite[Section IV]{haesaert2020robust} or \cite[Section IV]{huijgevoort2022similarity} respectively.

In \syscore, the construction of the finite-state abstraction is implemented in the functions \texttt{GridInputSpace} and \texttt{FSabstraction}. 
The function  \texttt{GridInputSpace} constructs the abstract input space \texttt{uhat}	by selecting a finite number of inputs from the input space \texttt{sysLTI.U}.
\begin{lstlisting}[numbers=none]
% Construct abstract input space
[uhat, InputSpace] = GridInputSpace(lu, sys.U, options);
\end{lstlisting} Here, \texttt{lu} is the number of abstract inputs in each direction and \texttt{options} are used to select an interface function from \eqref{eq:interface}. If interface function \eqref{eq:int2} or \eqref{eq:int_MOR} is chosen, \texttt{GridInputSpace} also divides the continuous input space into a part for actuation and for feedback, and returns these spaces as output \texttt{InputSpace}. This is done to make sure that the input bounds $u \in \mathbb{U}$ of the original model are satisfied. 
Next, we use  \texttt{FSabstraction} to compute a \emph{probability matrix} that contains the transition probabilities between states for all possible inputs in \texttt{uhat}.
\begin{lstlisting}[numbers=none]
% Construct abstract model
sysAbs = FSabstraction(sys, uhat, l, tol, DFA, options);
\end{lstlisting} Here, \texttt{sys} is the continuous-state system, 
\texttt{uhat} is the abstract input space $\mathbb{\hat{U}}$, \texttt{l} is the number of grid cells in each direction and \texttt{tol} is the tolerance for truncating to zero. This means that if a probability 
is smaller than the value set by \texttt{tol}, then we set it to zero to increase sparsity and hence decrease computation time. 
 Via efficient tensor computations, we split the computation of the probability matrix into two parts: one for the deterministic part of the transitions computed as a sparse matrix, and one for the stochastic part of the transitions. This reduces the required memory allocation and computation time drastically. For development purposes  \texttt{options} can be used to select whether or not to use  this efficient \emph{tensor} computation.
The complete probability matrix can then be obtained by using a tensor multiplication, however, we do not store the complete probability matrix and compute it when necessary in order to save memory.
\subsubsection*{Running example cont'd}
\begin{Example} To construct a finite-state abstraction of the car park model \texttt{sysLTI} (defined in code lines 1-13), we compute the abstract input space \texttt{uhat}: 
\end{Example}
\begin{fancyequations}
\begin{ExampleCode}
% Construct abstract input space uhat
lu = 3;  % number of abstract inputs
uhat = GridInputSpace(lu, sysLTI.U);
\end{ExampleCode}
\end{fancyequations}
\begin{Example}
and construct the abstract model \texttt{sysAbs} using the \texttt{DFA} constructed in code line 17 as follows:
\end{Example}
\begin{fancyequations}
\begin{ExampleCode}
% Construct finite-state abstraction
l = [200, 200];  % number of grid cells
tol = 10^-6;
sysAbs = FSabstraction(sysLTI, uhat, l, tol, DFA, 'TensorComputation', true);
\end{ExampleCode}
\end{fancyequations}

\subsection{Similarity quantification}\label{sec:simQuant}
To quantify the similarity between the model and its abstraction (either reduced order or finite state), we compute $\epsilon$ and $\delta$ such that they satisfy the $(\epsilon,\delta)$-stochastic simulation relation as defined in \cite[Definition 4]{huijgevoort2022similarity}. 
Here, $\epsilon$ and $\delta$ represent bounds on the output and probability deviations, respectively. This simulation relation allows us to consider scLTL specifications with unbounded time properties \cite{haesaert2020robust}.

When using model-order reduction, we construct two simulation relations, one relation $\mathscr{R}_{MOR}$ between the original model $M$ \eqref{eq:modelLTI} and reduced-order model $M_r$ \eqref{eq:modelRed}, and one relation $\mathscr{R}$ between $M_r$ and the finite-state model $\hat{M}$. The simulation relations are of the form
\begin{subequations}\label{eq:relations}
\begin{align} \SwapAboveDisplaySkip
		\mathscr{R}_{MOR} := \left\{ (x_r,x) \in  \mathbb{X}_r \times \mathbb{X} \mid ||x-Px_r||_{D_r} \leq \epsilon_r \right\}
	\label{eq:Rmor} \\
		\mathscr{R} := \left\{(\hat{x},x_r) \in \mathbb{\hat{X}} \times \mathbb{X}_r \mid ||x_r-\hat{x}||_D \leq \epsilon \right\}, \label{eq:simRel}
\end{align}
\end{subequations} 
with $||x||_D = \sqrt{x^\top D x}$ the weighted two-norm, where $D = D^\top \succeq 0$ is positive semi-definite. Following \cite{huijgevoort2022similarity}, these simulation relations can be combined into one total simulation relation between $M$ and $\hat{M}$. Following \cite[Section IV.A]{haesaert2020robust}, we can now compute the initial state of the reduced-order model as the state $x_{r,0}$ that minimizes $||x_0 - Px_{r,0} || _{D_r}$, that is 
 $x_{r,0} := (P^\top D_r P )^{-1}  P^\top D_rx_0$.

The computation of the simulation relation relies heavily on the coupling of the inputs $u,\hat{u}$ and disturbances  $w,\hat{w}$ of the two models. The inputs are coupled through an interface function and the disturbances via a coupling kernel. This is illustrated in Figure~\ref{fig:CouplingTotal} and is based on the method developed in \cite{huijgevoort2022similarity}. 
More specifically, the underlying computation is based on finding an invariant set for the error dynamics $x_{r,t+1} - \hat{x}_{t+1}$. To this end, an optimization problem constrained by parameterized linear matrix inequalities is used to find a value for $\delta$ that corresponds with the given value of $\epsilon$ \cite{huijgevoort2022similarity}. To solve this optimization problem, we use the \texttt{multi-parametric toolbox (MPT3)} \cite{herceg2013multi} with \texttt{YALMIP} \cite{lofberg2004yalmip} and with either solver  \texttt{SeDuMi} \cite{labit2002sedumi} or \texttt{MOSEK} \cite{aps2019mosek}.

In \syscore, similarity quantification is implemented in the function \texttt{QuantifySim}.
 \begin{lstlisting}[numbers=none]
% Quantify similarity
[simRel, interface] = QuantifySim(sys, sysAbs, epsilon, options)
\end{lstlisting}
This function quantifies the similarity between the models \texttt{sys} and \texttt{sysAbs}, with \texttt{sysAbs} either a reduced-order or a finite-state approximation of \texttt{sys}, hence in terms of behavior $\texttt{sysAbs} \preceq \texttt{sys}$.
\texttt{QuantifySim} yields a simulation relation \texttt{simRel} of the form \eqref{eq:relations} that is stored in the object \texttt{SimRel}. This object includes a method to check whether two states belong to the simulation relation and a method to combine the two simulation relations from \eqref{eq:relations} if necessary. Besides that, the function  \texttt{QuantifySim} also returns the feedback-matrix of the \texttt{interface} function, when interface \eqref{eq:int2} or \eqref{eq:int_MOR} is chosen through the  \texttt{options}.

\subsubsection*{Running example cont'd}
\begin{Example}
Next, we quantify the similarity between the model of the car stored in  \texttt{sysLTI} and its finite-state abstraction \texttt{sysAbs} constructed in code line 24 by choosing a suitable value for $\epsilon$ and using the function \texttt{QuantifySim}.
\end{Example}
\begin{fancyequations}
\begin{ExampleCode}
% Choose a value for epsilon
epsilon = 1.005;
% Quantify similarity
simRel = QuantifySim(sysLTI, sysAbs, epsilon);
\end{ExampleCode}
\end{fancyequations}

\smallskip\noindent \textbf{Piecewise affine systems.}
The function \texttt{QuantifySim} can handle both PWA \eqref{eq:modelPWA} and LTI models \eqref{eq:modelLTI}. However, for PWA systems the probability deviation is a PWA function $\boldsymbol{\delta}(\hat{x})$ that depends on the partition of the abstract state \cite{huijgevoort2022piecewiseaffineabstraction}.

\subsection{Synthesizing a robust controller}
We synthesize a robust (finite-state) controller based on the dynamic programming approach described in \cite{haesaert2020robust}, which is robust in the sense that it takes the deviation bounds $\epsilon$ and $\delta$ into account to compute a lower bound on the actual satisfaction probability. Furthermore, it is proven in \cite[Theorem 4]{haesaert2020robust} that the resulting control policy synthesized for the abstract model can always be refined to a control policy for the actual model. 

More specifically, we implicitly construct a product composition 
of the finite-state model $\hat{M}$ with the DFA such that computing the \emph{satisfaction probability} becomes a reachability problem over this product composition. 
This can in turn be solved using \emph{dynamic programming} by 
associating a robust dynamics programming operator that allows for an iterative computation of the lower bound on the satisfaction probability. Denote the state of the DFA by $q$, then the probability that a trajectory starting at $(\hat{x},q)$ reaches the set of accepting states by applying policy $\boldsymbol{\mu}$ within horizon $[1, 2, \dots N]$ is denoted as $V_N^{\boldsymbol{\mu}}(\hat{x},q)$. This is equivalent to the probability of satisfying the specification $\phi$ over this time horizon. The probability $V$ is computed iteratively by defining the operator
\begin{equation}
	\Bel^{\hat{u}} (V)(\hat{x},q) := \boldsymbol{L}\left( \mathbb{E}_{\hat{u}} \left( \min_{q^+ \in Q^+} \max\{1_{Q_f}(q^+), V(\hat{x}^+,q^+) \} \right) -\delta \right),
	\label{eq:BelOp}
\end{equation}
where $\hat x^+$and $q^+$ are resp. the next state of the abstract model and of the DFA, $\mathbb{E}$ is expectation with respect to the probabilistic transitions in the abstract model, $1_{Q_f}(q)$ is an indicator function that is equal to 1 if $q$ is inside the set of accepting states $Q_f$ of the DFA and is $0$ otherwise,
$\boldsymbol{L}: \mathbb{R} \rightarrow [0,1]$ is a truncation function, and with
\begin{equation} \label{eq:Qplus}
	Q^+(q,\hat{y}^+) := \left\{ \tau_{\mathcal{A}_\phi}(q,L(y^+)) \mid ||y^+-\hat{y}^+|| \leq \epsilon \right\},
\end{equation}
where $\tau_{\mathcal{A}_\phi}$ is the transition function of the DFA and $L(y^+)$ is the label of the next output. This operator is robust in the sense that the probability gets reduced by $\delta$ at every time step and the worst case transition of the DFA is considered with respect to $\epsilon$. The derivation of this operator for Markov decision processes can be found in \cite{haesaert2020robust}.

Synthesis of an abstract control strategy \texttt{pol} and the computation of the robust \emph{satisfaction probability} \texttt{satProb} is performed by the function \texttt{SynthesizeRobustController} and it is based on the abstract model \texttt{sysAbs}, the specification as a \texttt{DFA} and the 
simulation relation \texttt{simRel}.
\begin{lstlisting}[numbers=none]
% Compute satisfaction probabilty and policy
[satProb, pol] = SynthesizeRobustController(... 
sysAbs, DFA, simRel, thold, options)
\end{lstlisting}
We include the possibility to set the threshold \texttt{thold} that stops the value iteration when the difference between two iterations is smaller than this threshold. The default value is set to $1\cdot 10^{-12}$. This choice is justified by the fact that the operator in \eqref{eq:BelOp} is contractive and will always converge monotonically to a fixed-point.
Additionally, we include the \texttt{options} to compute the value function only for the initial DFA state and to compute an upper bound on the satisfaction probability. Internally, the dynamic programming algorithm computes the product between large-scale matrices (one of which is the probability matrix as mentioned in Section \ref{sec:abstr} on finite-state abstractions).
By performing these computations using a tensor product \cite{nilsson2018toward}, we gain superior computational efficiency.

The resulting control policy \texttt{pol}
is a mapping $\mu: \mathbb{\hat{X}} \times Q \rightarrow \mathbb{\hat{U}}$ from the pair of abstract and DFA states to the abstract input space. The abstract controller can now be written as $\hat{C}: \hat{u} = \mu(\hat{x},q)$.

 \subsubsection*{Running example cont'd}
 \begin{Example}
 After specifying the desired threshold for convergence \texttt{thold}, we synthesize a robust control policy \texttt{pol} based on the finite-state abstract model \texttt{sysAbs}, the specification as a \texttt{DFA} and the simulation relation \texttt{simRel} constructed in code line 28. In this case, we are only interested in the satisfaction probability \texttt{satProb} of the initial DFA state, hence we set the \texttt{options} to \texttt{true}.
\end{Example}
\begin{fancyequations}
\begin{ExampleCode}
% Specify threshold for convergence error
thold = 1e-6;
% Synthesize an abstract robust controller
[satProb, pol] = SynthesizeRobustController(...
sysAbs, DFA, simRel, thold, true);
\end{ExampleCode}
\end{fancyequations}
\begin{Example}
	The robust satisfaction probability is computed for all $x_0 \in \mathbb{X}$. For initial states $x_0 = [-4,-5]^\top$, $x_0 = [-8, 2]^\top$, and $x_0 = [4, 8]^\top$, it equals respectively $0.60, 0.52$, and $0.42$.
\end{Example}

\subsection{Control refinement}
To refine an abstract finite-state controller to a controller $C$ that can be implemented on the original continuous-state system (see step 5 in Figure~\ref{fig:TLcontrol}) we use one or multiple interface functions from \eqref{eq:interface} as illustrated in Figure~\ref{fig:CouplingTotal}. In \syscore, control refinement is included in the class \texttt{RefineController}, where it is possible to select an interface function using the \texttt{options}. 
\begin{lstlisting}[numbers=none]
% Refine abstract controller
Controller = RefineController(satProb, pol, sysAbs, simRel, sys, DFA, options);
\end{lstlisting} This class not only refines the finite-state input to the actual input, but also determines the state of the finite-state model based on the state of the original model. 

\subsubsection*{Running example cont'd}
\begin{Example}
	To construct a controller $C$ that can be implemented on the original model $M$ based on the abstract control policy \texttt{pol} computed in code line 32, we use the following.
\end{Example}
	\begin{fancyequations}
\begin{ExampleCode}
% Refine abstract controller 
Controller = RefineController(satProb, pol, sysAbs, simRel, sysLTI, DFA);
\end{ExampleCode}
	\end{fancyequations}

\subsection{Deployment}
The final step is to deploy the controller on the model and perform simulations using \texttt{ImplementController}.

\begin{lstlisting}[numbers=none]
% Implement the controller on the model
xsim = ImplementController(x0, N, Controller, option);
\end{lstlisting} Here, \texttt{N} is the desired time horizon for the simulation and \texttt{option} is used to supply the number of trajectories and/or additional model-order reduction inputs.

\subsubsection*{Running example cont'd}
\begin{figure}
	\centering
	\includegraphics[width=.75\columnwidth]{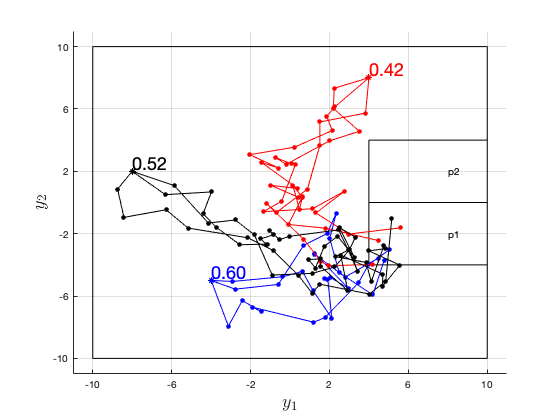}
	\caption{Trajectories for the running example. Three trajectories are obtained for each initial state: $x_0 = [-4, -5]^\top$ (blue), $x_0 = [-8, 2]^\top$ (black), and $x_0 = [4, 8]^\top$ (red). The corresponding robust satisfaction probability is given at the inital state.}
	\label{fig:RunExTraj}
\end{figure}

\begin{Example}
To simulate the controlled system with the \texttt{Controller} constructed in code line 34, we use \texttt{ImplementController} to obtain the state trajectory starting at $x_0$. Trajectories of the controlled system with three initial states are illustrated in Figure \ref{fig:RunExTraj}. 
\end{Example}
\begin{fancyequations}
\begin{ExampleCode}
x0 = [-4; -5]; % initial state
N = 40; % time horizon
% Simulate controlled system
xsim = ImplementController(x0, N, Controller);
\end{ExampleCode}
\end{fancyequations}

\section{Benchmarks}\label{sec:benchmarks}
To show the capabilities of \syscore, we included multiple benchmarks, of which some are discussed here. The package delivery has a complex specification with a cyclic DFA, the building automation system includes model-order reduction and the Van der Pol oscillator is nonlinear. We evaluate the run time and memory usage of the benchmarks, and compare \syscore to some existing tools. 

\subsection{Package delivery}
With the package delivery benchmark \cite{ARCH22}, we show the capability of \syscore to handle complex scLTL specifications beyond basic reach-avoid scenarios, i.e., cyclic DFAs. Consider an agent traversing in a 2D space, whose dynamics can be described by an LTI system \eqref{eq:modelLTI} with $A := 0.9I_2$, $B := I_2$, $B_w := \sqrt{0.2}I_2$, $C :=I_2$, and disturbance $w_k\sim\mathcal{N}(0, I_2)$. We initialize the system using \texttt{LinModel}.

Define the state space $\X = [-6,6]^2$, input space $\U = [-1,1]$, output space $\Y = \X$, and regions $p_1$, $p_2$ and $p_3$ as follows: $p_1 := [5, 6] \times [-1 , 1]$, $p_2 := [0, 1] \times [-5, 1]$ and $p_3 := [-4, -2] \times [-4, -3]$.
The agent can pick up a package at $p_1$ and must deliver it to $p_3$. If the agent visits $p_2$ while carrying a package, it loses the package and has to 
pick up a new package at $p_1$.
This corresponds to the scLTL specification $\lozenge(p_1 \wedge (\neg p_2 \until p_3))$ implemented as in \eqref{eq:phi_PD}. We generate the corresponding DFA using \texttt{TranslateSpec}.

\begin{figure*}[t]
	\centering
	\subfloat[Robust satisfaction probability of the package delivery benchmark.]{\includegraphics[width=0.3\textwidth]{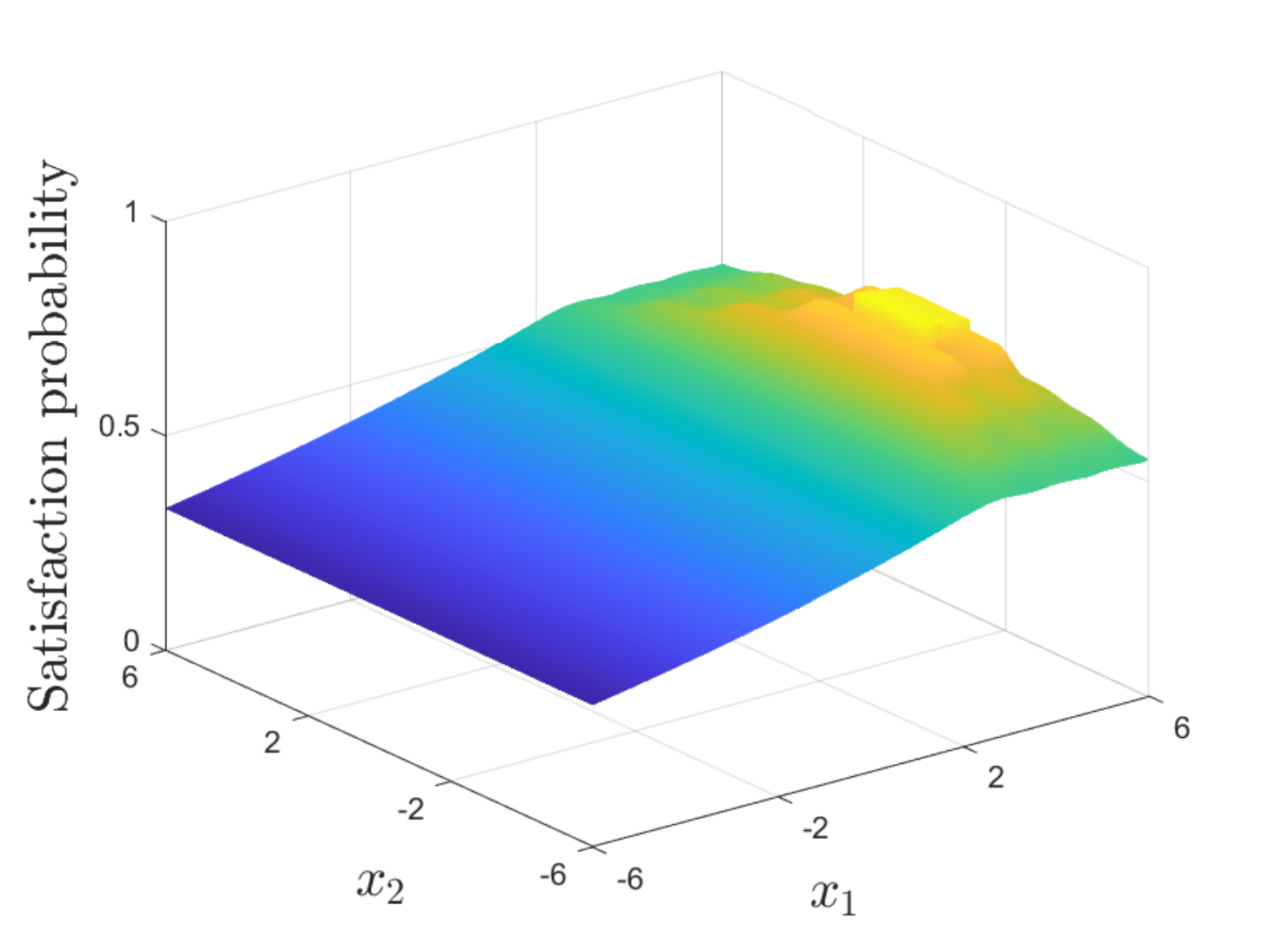}\label{fig:PckDelivery_SatProb}} \quad
	\subfloat[Robust satisfaction probability of the Van der Pol oscillator benchmark. ]{\includegraphics[width=0.3\textwidth]{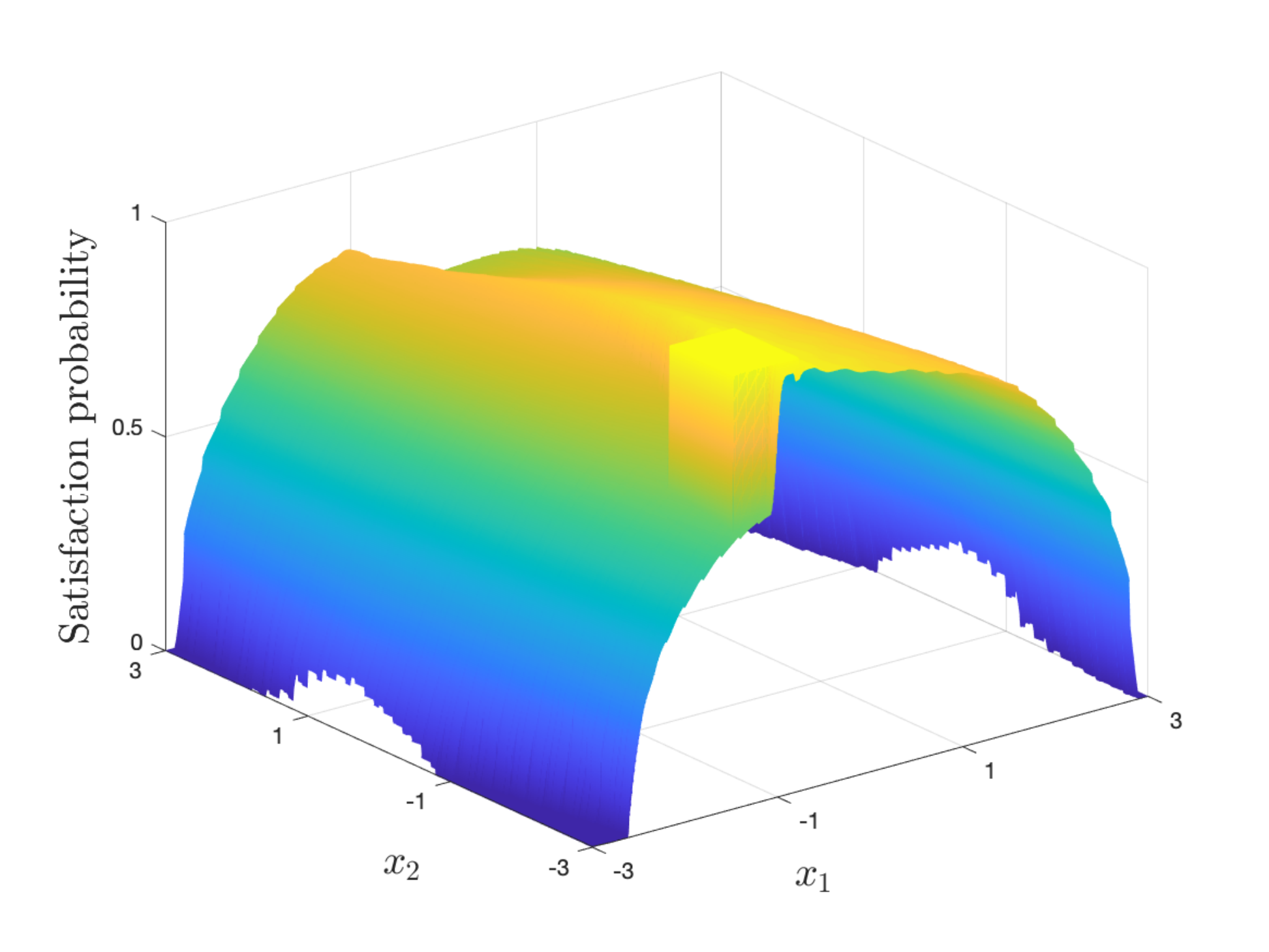}\label{fig:VdP_SatProb}} \quad 
	\subfloat[Robust satisfaction probability of the reduced-order model of the building automation system benchmark. Yellow and blue correspond to a probability of $0.9035$ and $0$ resp. ]{\includegraphics[width=0.3\textwidth]{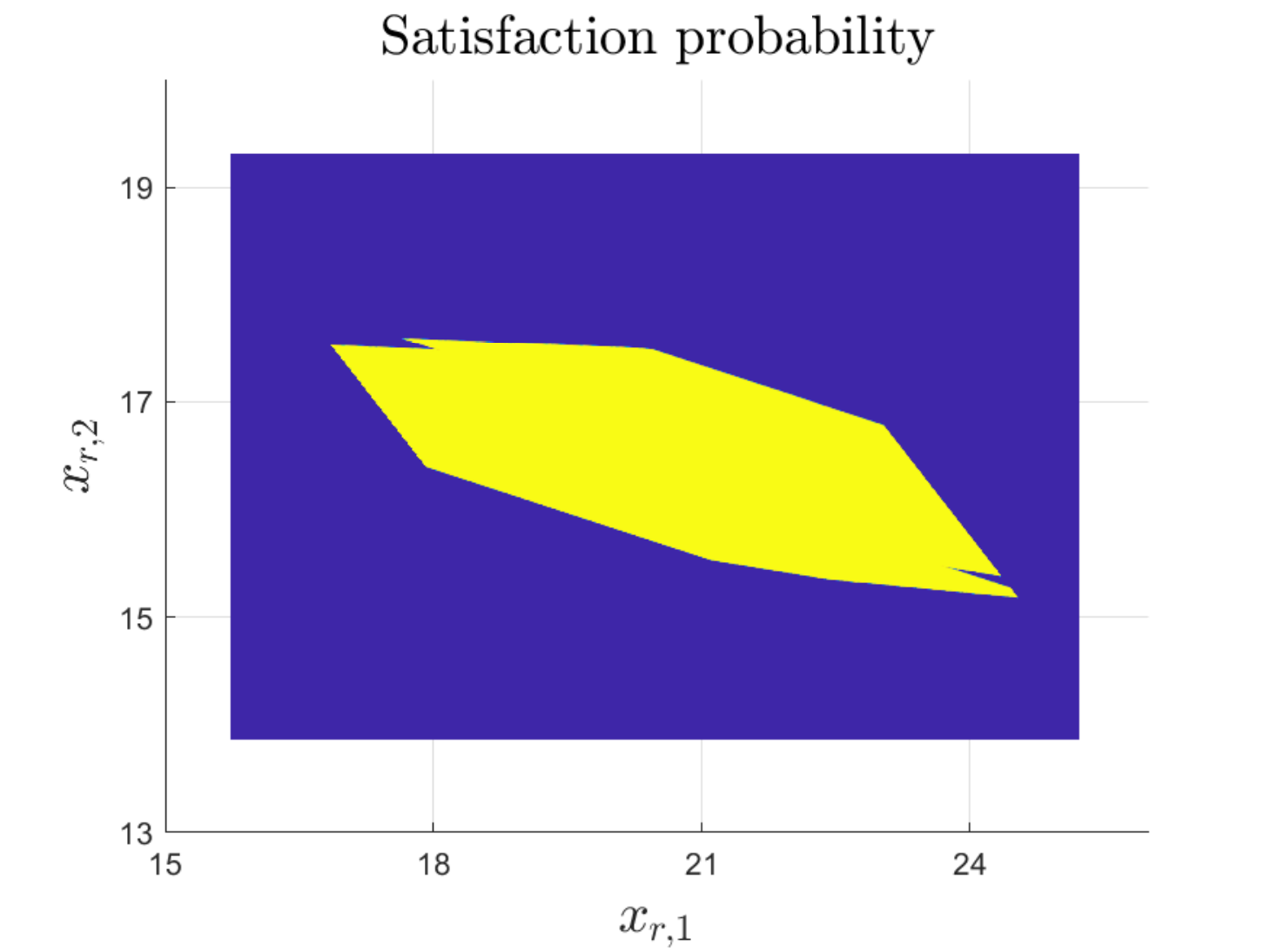}\label{fig:BAS_satProb}}
	\caption{Robust satisfaction probability of the initial DFA state as a function of the initial state for the different benchmarks. In (a) the package delivery benchmark, in (b) the van der Pol benchmark, and in (c) the building automation system.}
	\label{fig:SatProb_all}
\end{figure*}

Next, we construct a finite-state abstraction using \texttt{GridInputSpace} and \texttt{FSabstraction}. For this study, we choose a comparatively fine state abstraction $l = [400, 400]$, which allows us to generate a simulation relation using \texttt{QuantifySim} with an epsilon of just 0.075. Note that the partition size $l$ is a tuning parameter which is determined empirically.
We synthesize a robust controller for the discrete abstraction using
\begin{fancyequations}
\begin{ExampleCode}[numbers=none]
[satProb, pol] = SynthesizeRobustController( ...
	sysAbs, DFA, rel, thold, false);
\end{ExampleCode}
\end{fancyequations}
Since the resulting control policy is conditional on both the current system state and the DFA state, we set the 5th argument to \texttt{false}. By doing so, we synthesize a controller for all DFA states instead of only the initial one. The obtained robust satisfaction probability \texttt{satProb} over different initial states $x_0$ is displayed in Figure~\ref{fig:PckDelivery_SatProb} and can be obtained by running 
\begin{fancyequations}
\begin{ExampleCode}[numbers=none]
% Plot satisfaction probability
plotSatProb(satProb, sysAbs, 'initial', DFA);
\end{ExampleCode}
\end{fancyequations}
The peak satisfaction probability is 0.663.

Finally, we refine the controller using \texttt{RefineController}.
To demonstrate the performance of the obtained controller, we simulate the controlled system using \texttt{ImplementController} for $N=60$ time steps and an initial state of $x_0=[-5,-5]^T$. 
Note that $N$ is an empirical parameter and should be set high enough for the DFA to terminate.
As expected, the agent moves to region $p_1$ to pick up a package, and delivers it to $p_3$ whilst avoiding $p_2$. To plot trajectories we included the function \texttt{plotTrajectories}.

\subsection{Van der Pol oscillator}
In this benchmark, we show how \syscore can be applied to nonlinear stochastic systems. For this, consider the discrete-time dynamics of the Van der Pol oscillator \cite{ARCH22}, given by
\begin{align}\SwapAboveDisplaySkip
	x_{1,t+1} &= x_{1,t} + x_{2,t} \tau+w_{1,t}
	\label{eq:vdp}
	\\
	x_{2,t+1} & = x_{2,t} + (-x_{1,t}+(1-x_{1,t}^2)x_{2,t}) \tau+ u_t+w_{2,t},
	\nonumber
\end{align}
where the sampling time $\tau$ is set to $0.1s$, $w_t\sim\mathcal{N}(0,0.2I_2)$, and $y_t=x_t$.
We define the state space $\X = [-4,4]^2$, input space $\U = [-1,1]$, and output space $\Y = \X$. 
For the Van der Pol oscillator, we are looking at an unbounded safety specification (cf. \eqref{eq:phi_VdP}), where the objective is to synthesize a controller such that the system remains in the region $p_1 := \X$ until reaching region $p_2 := [-1.4,-0.7] \times [-2.9, -2]$, corresponding to the scLTL specification $p_1 \until p_2$. First, we construct a DFA for the formula \eqref{eq:phi_VdP} using \texttt{TranslateSpec}.

Since the dynamics of the oscillator (\texttt{sysNonLin}) in \eqref{eq:vdp} are nonlinear, the abstraction process is split into two parts as outlined in Section~\ref{sec:abstr}. First, we construct a PWA approximation as follows:
\begin{fancyequations}
\begin{ExampleCode}[numbers=none]
% Number of grid points in each direction
N = [41 41]; 
% Perform PWA approximation
sysPWA = PWAapproximation(sysNonLin, N);
\end{ExampleCode}
\end{fancyequations}

In the second part of the abstraction step, a finite-state abstraction (\texttt{sysAbs}) of the PWA approximation (\texttt{sysPWA}) is constructed using \texttt{GridInputSpace} and \texttt{FSAbstraction} with \texttt{l=[600,600]} grid cells. To generate a simulation relation between this abstraction and the original model, we set $\epsilon=0.1$ and compute a suitable weighting matrix $D$ for the simulation relation on $(\hat{x},x)$, as described in Section~\ref{sec:simQuant}. To reduce computation time, we only use a finite number of states to compute this weighting matrix. Details on why we need this global weighting matrix can be found in \cite{huijgevoort2022piecewiseaffineabstraction}. 
\begin{fancyequations}
\begin{ExampleCode}[numbers=none]
% Compute weighting matrix D for the simulation relation based on the following states
States = [1/8*x1l, 6/10*x2u; 5/7*x1u, 5/17*x2u; 2/13*x1u, 5/9*x2l; 3/4*x1l, 1/7*x2l; 0, 0]';
[D, ~] = ComputeD(epsilon, sysPWA, sysAbs, 'interface', int_f, 'states', States);
% Quantify similarity
[rel, sysPWA] = QuantifySim(sysPWA, sysAbs, epsilon, 'interface', int_f, 'weighting', D);
\end{ExampleCode}
\end{fancyequations}

Note that \texttt{QuantifySim} returns \texttt{sysPWA} instead of the usual \texttt{interface}, because each piecewise-affine system gets its own interface function and we store this directly in \texttt{sysPWA}.

Next, we use \texttt{SynthesizeRobustController} to synthesize a robust controller for \texttt{sysAbs} and show the satisfaction probability (displayed in Figure~\ref{fig:VdP_SatProb}) using \texttt{plotSatProb}. Finally, we refine the controller as follows:
\begin{fancyequations}
\begin{ExampleCode}[numbers=none]
Controller = RefineController(satProb, pol, sysAbs, rel, sysPWA, DFA, int_f);
\end{ExampleCode}
\end{fancyequations}
As before, \texttt{ImplementController} is used to simulate the system.
	\begin{table*} \caption{An overview of the different benchmarks and their total computation time in seconds (s) and memory usage in megabyte (MB). The details of the computation times for each step are reported in Table~\ref{tab:compTime}. \emph{Dim.} and \emph{Comp.} are abbreviations for \emph{Dimension} and \emph{Computation}, respectively. The size of the specification refers to the number of states of the DFA.}
		\label{tab:Tutorials}
		\begin{tabular}{| c | c c | c | c  c c | p{2cm} | p{1.8cm} |} \hline 
			Benchmark 						& \multicolumn{2}{c|}{System} & MOR & \multicolumn{3}{c|}{Specification} & Comp. time (s)	& Memory (MB)\\  
			& 	Dynamics				& Dim. 			& & Type & Time horizon  & Size					 	& 											&  \\ \hline
			Running example 	& Linear	& 2	& 	No		& Reach-avoid & Unbounded		& 3	& 7.94	& 27.53\\
			Package delivery	& Linear	 & 2	& No 		& Reach-avoid	& Unbounded		& 3	& 11.02		& 133.4 \\
			Van der Pol oscillator & Nonlinear & 2 & No 	& Safety, reachability & Unbounded & 3 & 3191.6 & 178.83 \\
			Building automation	&	Linear		& 7 & Yes		& Safety			& Bounded				& 8 & 122.05 & 5365.6 \\
			\hline 
		\end{tabular}
	\end{table*}
\begin{table*} \caption{Computation times for the different steps (1)-(6) in seconds and as percentage of the total runtime. Steps (1)-(6) correspond to (1) translating the specification, (2) finite-state abstraction, (3) similarity quantification, (4) synthesizing a controller, (5) control refinement, and (6) deployment. Step (5) is almost instantaneous ($\approx$ 0.001 s), therefore, we have taken the computation times of steps (5) and (6) together.} \label{tab:compTime}
	\begin{tabular}{| c | c | c | c | c | c | c |  } \hline 
		& Step (1) & Step (2) & Step (3)  & Step (4) & Step (5) and (6) & Total \\  \hline 
		Running example & 0.259s	(3.26\%) 	& 1.316s (16.57\%) 	& 5.590s (70.38\%) & 0.507s (6.39\%)	& 0.204s (2.57\%) & 7.944s (100\%)\\
		Package delivery & 	0.284s (2.58\%)	& 1.657s (15.04\%) 	& 6.193s (56.2\%) 	& 1.708s (15.5\%) 	& 0.702s (6.37 \%)	& 11.02s (100\%)    \\
		Van der Pol oscillator & 0.590s (0.02\%)  & 1440.1s (45.1\%) 	& 1748.6s (54.8\%) & 2.854s (0.09\%)  & 1.417s (0.04\%) & 3191.6s (100\%) \\ 
		Building automation & 0.361s (0.30\%)	& 4.80s (3.94\%)	& 	67.92s (55.7\%)	& 37.33s (30.6\%)	& 9.19s (7.53\%)  & 122.05s (100\%)  \\
		\hline 
	\end{tabular}
\end{table*}
\subsection{Building automation system}
In the last benchmark, we address a large-scale system showcasing the model-order reduction capabilities of \syscore. We consider a 7D \emph{affine} stochastic system of a building automation system, regulating the temperature in two zones influenced by a 6D disturbance. A detailed description including the system dynamics can be found in \cite{ARCH18,cauchi2018benchmarks}.
The goal is to synthesize a controller maintaining the temperature in zone one at 
$20^{\circ}C$ with a maximum permissible  
deviation of $\pm0.5^{\circ}C$ for 6 consecutive time steps. We translate the specification \eqref{eq:phi_BAS} to a DFA using \texttt{TranslateSpec}.

The dynamics of this building automation system are not of the form \eqref{eq:modelLTI}, since it is influenced by a Gaussian disturbance with mean $\mu \neq 0$ and variance $\Sigma \neq I$. Furthermore, it is not an LTI system, but affine, which cannot be handled by our current implementation of model-order reduction. To deal with the disturbance, we first transform the system to a system with Gaussian disturbance $w \sim \mathcal{N}(0,I)$ using the following:
\begin{fancyequations}
\begin{ExampleCode}[numbers=none]
% Transform the model
[sysLTI, a] = NormalizeDisturbance(sysLTI,a);
\end{ExampleCode}
\end{fancyequations}

To deal with the affine dynamics, we perform a steady-state shift and simulate the steady-state system that has LTI dynamics. After performing the control synthesis steps, we compensate for this steady-state shift again to obtain the dynamics of the actual system. 

Now, we can start with the synthesis steps. First, we reduce the 7D model to a 2D reduced-order model (see Eq.~\eqref{eq:modelRed})
using function \texttt{ModelReduction} with \texttt{f = 0.098} and \text{dimr=2}.
\begin{fancyequations}
\begin{ExampleCode}[numbers=none]
% Perform model-order reduction
[sysLTIr, ~] = ModelReduction(sysLTI, dimr, f);
\end{ExampleCode}
\end{fancyequations}

As mentioned in Section~\ref{sec:mor}, we use an interface function of the form \eqref{eq:int_MOR}, which is selected using \texttt{int\_f = 1} and %initialized by 
compute the matrices $P$ and $Q$ using \texttt{ComputeProjection}.
Next, we define the state and input spaces, and the output regions and APs for the reduced-order model as before.

To construct the finite-state abstraction of the reduced-order model, we first grid the input space with \texttt{lu = 3}.
\begin{fancyequations}
\begin{ExampleCode}[numbers=none]
% Construct abstract input space
[uhat,sysLTIr.U] = GridInputSpace(lu, sysLTIr.U, 'interface', int_f, 0.6, 0.175);
\end{ExampleCode}
\end{fancyequations} Here, we have chosen to use $60\%$ of the input space for actuation and $17.5 \%$ for feedback. This leaves $22.5 \%$ for the $Qx_{r,t}$ part of the interface function, which is currently not guaranteed to be satisfied.

Before constructing a finite-state abstraction of the reduced-order model, we reduce the state space to increase the computational speed. This step is currently only available for invariance specifications and is performed by \texttt{ReduceX}, which  performs a number of backwards iterations on the safety region $P_1$ to determine a good guess of the invariant set. This set is then used as the reduced state space. The construction of the finite-state abstraction of the reduced-order model is as before, except that we give the total number of grid cells as input \texttt{l}, instead of the number of grid cells in each direction. 
\begin{fancyequations}
\begin{ExampleCode}[numbers=none]
% Reduce the state space to speed up computations
[sysLTIr, ~] = ReduceX(sysLTIr, sysLTIr.U{2}, P1, 'invariance', 5);
% Construct finite-state abstraction
l = [3000*3000];  % Total number of grid cells
tol=10^-6;
sysAbs = FSabstraction(sysLTIr, uhat, l, tol, DFA, 'TensorComputation', true);
\end{ExampleCode}
\end{fancyequations}

To relate the reduced-order finite-state model \texttt{sysAbs} to the original model \texttt{sysLTI}, we construct two simulation relations: relation \texttt{rel\_1} with $\epsilon_1=0.2413$ between \texttt{sysLTI} and \texttt{sysLTIr}, and relation \texttt{rel\_2} with $\epsilon_2 = 0.1087$ between \texttt{sysLTIr} and \texttt{sysAbs}:
\begin{fancyequations}
\begin{ExampleCode}[numbers=none]
% Compute MOR simulation relation
[rel_1, K, kernel] = QuantifySim(sysLTI, sysLTIr, epsilon_1, 'MOR', sysAbs);
% Compute finite-state simulation relation
[rel_2] = QuantifySim(sysLTIr, sysAbs, epsilon_2);
% Combine simulation relations
rel = CombineSimRel(rel_1, rel_2, sysLTIr, sysAbs);
\end{ExampleCode}
\end{fancyequations}
For model-order reduction we have to explicitly define the coupling \texttt{kernel} matrix $F$, that is later used to compute the disturbance of the reduced-order model as $w_r\! =\! w+ F(x\!-\!Px_r)$.  For details see \cite{huijgevoort2022similarity}.

Synthesizing and refining the controller are done as before and the satisfaction probability of the reduced-order model is shown in Figure~\ref{fig:BAS_satProb} (obtained through \texttt{plotSatProb}).
We simulate the controlled system $N_s=6$ times, making sure the output is shifted with respect to the steady-state solution. 
\begin{fancyequations}
\begin{ExampleCode}[numbers=none]
% Simulate controlled system Ns times
N_s = 6;
xsim = ImplementController(x0, N, Controller, Ns, 'MOR', sysLTIr, kernel);
\end{ExampleCode}
\end{fancyequations}
The resulting trajectories can be evaluated using \texttt{plotTrajectories}.
\begin{table*}
	\caption{Results of the benchmarks for different tools.
		Here, \emph{n.a.} means that a tool is \emph{not applicable} and \emph{n.s.} means that the current version of the tool does \emph{not natively support} the computations on the benchmark, but that we do not see fundamental limitations hindering such an extension. 
		To compare the tools we exclude the deployment of the controller (step (6)), since this step is not performed by the other tools.
	}
	\label{tab:ToolComp}
	\subfloat[Package delivery benchmark.]{
		\begin{tabular}{|c | c | } 
			\hline
			Tool & Run time  (s)  \\
			\hline
			\amytiss & n.s.  \\ 
			\faust & n.s.  \\ 
			\sreachtools &  n.a. \\ 
			\stochy & n.s.  \\ 
			\syscore & 10.319  \\ \hline 
		\end{tabular}
		\label{tab:PDcomp}
	}
	\qquad 
	\subfloat[Van der Pol benchmark.]{
		\begin{tabular}{|c | c |} 
			\hline
			Tool & Run time  (s)  \\ 
			\hline
			\amytiss & n.s.  \\ 
			\faust & n.a.  \\ 
			\sreachtools &  n.a.  \\ 
			\stochy & n.a.  \\ 
			\syscore & 3190.2  \\ \hline 
		\end{tabular}
		\label{tab:Polcomp}
	}
	\qquad 
	\subfloat[Building automation benchmark.]{
		\begin{tabular}{|c | c | c |} 
			\hline
			Tool & Run time  (s) & Max. reach probability \\ 
			\hline
			\amytiss & 312.14 & $\approx 0.8$ \\ 
			\faust & n.s. & n.s. \\ 
			\sreachtools &  4.59 & $\geq 0.99$ \\ 
			\stochy & $\geq$ 335.876 & $\geq 0.8 \pm 0.23$ \\ 
			\syscore & 112.86 & $\geq 0.9035$ \\ \hline 
		\end{tabular}
		\label{tab:BAScomp}
	}
\end{table*}
\subsection{Performance evaluation}\label{sec:perform}
The performance of \syscore is evaluated on the benchmarks mentioned above. The details of the benchmarks and their total run time and memory usage are reported in Table~\ref{tab:Tutorials}. The computation times per step are reported in Table~\ref{tab:compTime}. The data has been obtained on a computer with a 2,3 GHz Quad-Core Intel Core i5 processor and 16 GB 2133 MHz memory by taking the average over $5$ computations. Here, we observed a maximum $6\%$ standard deviation.

Table~\ref{tab:Tutorials} can be used to compare the different benchmarks with respect to the computations performed by \syscore. The main difference between the running example and the package delivery benchmark is the DFA. The DFA of the package delivery benchmark requires more memory, however, the increase in computation time is small. Due to the simple DFA of the running example, we only compute the satisfaction probability for the initial DFA state.
This will not suffice for the package delivery benchmark, which is the reason that more computation time is spent on steps (4)-(6) compared to the running example (see Table~\ref{tab:compTime}).
The computation time for the nonlinear benchmark is large, however, the memory usage remains reasonable. The increase in computation time is mainly due to the fine gridding. We can also see in Table~\ref{tab:compTime} that the similarity quantification takes a considerable amount of time. This is because we perform this step for each partition separately (1600 times in this case). 
For higher-dimensional systems that require model-order reduction (building automation system benchmark), the computation time and memory usage increase substantially, mainly due to the the fact that the similarity quantification has to be performed multiple times. However, we also see from Table~\ref{tab:compTime} a large increase in the computation time for the controller synthesis.

Table~\ref{tab:compTime} shows that the similarity quantification of step (3) requires the most computation time, followed by the finite-state abstraction of step (2). 
The large computation time of the similarity quantification is due to solving an optimization problem constrained by parameterized matrix inequalities that could be non-convex.
For most abstraction-based approaches in the literature, the main bottleneck is the finite-state abstraction. This shows the efficiency of our tensor-based implementations. It should also be noted that the efficiency of the tensor computations is also exploited in the control synthesis step.

\subsection{Comparison to existing tools}
A comparison of the results on the benchmarks obtained by \syscore and current tools is given in Table~\ref{tab:ToolComp}.
The package delivery benchmark has a complex DFA and cannot be handled natively by tools \amytiss, \faust, and \stochy (see Table~\ref{tab:PDcomp}).
\sreachtools can only handle safety specifications and is not applicable to this benchmark.
The Van der Pol oscillator benchmark poses significant challenges for the tools due to its nonlinear dynamics, as reported in Table~\ref{tab:Polcomp}. Only \amytiss can solve a benchmark that resembles this one as considered in \cite{ARCH21} with multiplicative noise instead of additive noise. \amytiss can only handle systems with a bounded disturbance, hence it cannot directly solve the benchmark as presented here.

The benchmark on the building automation system can be solved by \amytiss, \sreachtools, and \stochy without being able to use model-order reduction. This benchmark considers a stochastic safety problem and the performance of multiple tools is compared in \cite{ARCH20,ARCH21}. 
Table~\ref{tab:BAScomp} reports the results of \syscore together with the results from running the repeatability packages of \cite{ARCH20, ARCH21} on a computer with a 2,3 GHz Quad-Core Intel Core i5 processor and 16 GB 2133 MHz memory. For \stochy, there was no repeatability package available, however, since the computational power of the CPU used for the results in \cite{ARCH21} was more than our computer, we included the results of \cite{ARCH21} as a lower bound on the computation time required by \stochy.
Note that this benchmark belongs to the class of partially degenerate systems \cite{soudjani2013probabilistic}. The formulation of the abstraction error for this class is available but the current version of \faust does not natively support partially degenerate systems.
With respect to the computation time,  \sreachtools performs best, and \amytiss and \stochy require a longer computation time.
Though from the results in  \cite{ARCH21}, we see that \amytiss could be faster than the current implementation of \syscore when parallel execution within CPUs is available (this parallel computation will be exploited in future versions of \syscore).
With respect to accuracy, both \amytiss and \stochy obtain a maximum reachability probability smaller than \syscore, while \sreachtools still outperforms \syscore. This shows that \sreachtools is the best option for this benchmark, which is expected since it is developed exactly for linear systems and stochastic reach-avoid problems with small disturbances. 

\section{Summary and extensions}
This paper described the first release of \syscore, a tool that excels at control synthesis problems for systems with a large (unbounded) stochastic disturbances and temporal specifications with possibly unbounded time horizon. It combines reduced-order models and finite abstractions with formal guarantees obtained by coupled stochastic simulation relations.  
\syscore substantially extends the class of models and temporal specifications that current tools can handle for control synthesis. 
Furthermore, the modular development of \syscore allows ease of use and facilitates future extensions. 
The efficient implementation of tensor computations in \syscore allows for fast computations, which can be exploited further by including more parallel computations as done in \amytiss.

An important direction for future releases is to extend the current implementation of model-order reduction to piecewise-affine systems, such that it can also be applied to nonlinear systems. 
Currently, only Gaussian disturbances are implemented in \syscore, however, extensions to other distributions are under way and require deriving new inequality constraints for the optimization problem solved in the similarity quantification.
The computation time of the similarity quantification is large due to solving optimization problems constrained by parameterized matrix inequalities that could be non-convex. We are working on improving the efficiency of solving this optimization.

The modular implementation of \syscore can be utilized to integrate model-order reduction with discretization-free approaches such as
the kernel method of \sreachtools \cite{vinod2019sreachtools,thorpe2021sreachtools} and the
 barrier certificates \cite{jagtap2020formal}, or to perform synthesis for stochastic systems with parametric uncertainty \cite{Schon2022subsim}.
To get non-trivial lower bounds, \syscore currently requires fine-tuning the hyper parameters (e.g., the grid size and the output deviation). It is of interest to automatically design these parameters or to provide guidelines to the user on the appropriate values depending on the case study.

\bibliographystyle{acm}
\bibliography{Sources}

\end{document}